\documentclass{PoS}
\usepackage{graphicx}
\usepackage{axodraw}
\usepackage{cite}

\title{Using $K^0\pi^-\to\pi^-$ transitions to compute
$K\to(\pi\pi)_{I=0}$ decay amplitudes at NLO in the chiral
expansion}

\ShortTitle{$K\pi\to\pi$ transitions}

\author{\speaker{Changhoan Kim} and Christopher Sachrajda\\
        School of Physics and Astronomy \\
        University of Southampton
        Highfield, Southampton, UK\\
        E-mail: \email{chateau@phys.soton.ac.uk}, \email{cts@phys.soton.ac.uk}}

\abstract{It is proposed to compute matrix elements for the
(unphysical) $K^0\pi^-\to \pi^-$ transition to determine the
next-to-leading order low energy constants of the weak chiral
Lagrangian. This allows us to evaluate $K\to(\pi\pi)_{I=0}$ decay
amplitudes at this level of precision. This approach has several
significant advantages over the use of $K\to\pi\pi$ transitions,
most notably the elimination of $s$-channel disconnected diagrams
and the use of fewer inversions.}

\FullConference{The XXV International Symposium on Lattice Field Theory\\
         July 30-4 August 2007\\
         Regensburg, Germany}

\newcommand{\p}{\partial}
\newcommand{\pslash}{p\kern-1ex /}
\newcommand{\lslash}{l\kern-1ex /}
\newcommand{\kslash}{k\kern-1ex /}
\newcommand{\dslash}{\p\kern-1.2ex /}
\newcommand{\Dslash}{{\cal D}\kern-1.5ex /}
\newcommand{\Aslash}{A\kern-1.2ex /}

\newcommand{\vev}[1]{\left\langle #1 \right\rangle}

\newcommand{\CPV}{CP\kern-16pt\rotatebox[x=0mm,y=1mm]{-36}{\Huge /}}

\newcommand{\cO}{{\cal O}}

\newcommand{\cH}{{\cal H}}

\begin{document}

\section{Introduction}
Nonleptonic kaon decays present a number of major challenges.
Direct $CP$ violation in kaon decays has been confirmed by the
non-zero measurements of $\varepsilon^\prime /
\varepsilon$\cite{Batley:2002gn,AlaviHarati:2002ye}, but a
quantitative theoretical understanding, even within the standard
model, is still lacking. The $\Delta I=1/2$ rule also remains a
longstanding puzzle. Whilst these are weak decays, the relevant
energy scale is of the order of a few hundred MeV at which
non-perturbative QCD effects are significant making the evaluation
of the amplitudes difficult.

The lattice formulation of QCD together with large scale numerical
simulations provide the opportunity to compute the
non-perturbative effects from first principles. However, with the
computing resources which are currently available it is not
possible to perform the simulations directly at the physical
values of $m_u$ and $m_d$. Although the situation is rapidly
improving, it is still unlikely that computations with pion masses
of about 140\,MeV will be performed in the near future. It is
proving particularly useful in general, and for kaon decays in
particular, to use chiral perturbation theory ($\chi$PT) to obtain
physical results from simulations with heavier values of
$m_{u,d}$~\cite{Bernard:1985wf}. In this approach, the QCD
operators in the effective weak Hamiltonian are written in terms
of a sum of operators composed of meson fields in the same chiral
representation. A priori, the coefficients of the operators in the
resulting Weak Chiral Lagrangian, i.e. the \textit{low energy
constants} (LECs), are unknown. The aim is to determine the LECs
from the mass and momentum dependence of matrix elements computed
in lattice simulations. This is well understood for $\Delta I=3/2$
$K\to\pi\pi$ decays in which the two-pion state has isospin $I=2$.
In this paper we focus on $\Delta I=1/2$ decays in which the two
pions have $I=0$

At leading order in $\chi$PT, there are only a small number of
LECs. For example, for the $\Delta S=1$ operator in the $(8,1)$
representation there are 2 LECs, $\alpha^{(8,1)}_1$ and
$\alpha^{(8,1)}_2$, in terms of which the $K\to\pi \pi$, $K\to\pi$
and $K\to$\,vacuum matrix elements are:
\begin{eqnarray}
\vev{\pi^+\pi^-|\cO^{(8,1)}|K^0}&=& \frac{4i}{f^3}(m_K^2 -
m_\pi^2) \alpha^{(8,1)}_1\,,\\
\vev{\pi^+| \cO^{(8,1)} |K^0 } &=& \frac{4m_M^2}{f^2}(
\alpha^{(8,1)}_1 - \alpha^{(8,1)}_2 )\,,\\
\vev{0| \cO^{(8,1)} |K^0 }& = &\frac{2i}{f}( m_K^2 - m_\pi^2)
\alpha^{(8,1)}_2\,.
\end{eqnarray}
Thus by computing the $K\to\pi$ and $K\to$\,vacuum matrix
elements, $\alpha^{(8,1)}_{1,2}$ can be determined and the
$K\to\pi\pi$ matrix elements can be evaluated at leading order.
Such calculations were performed in 2001 with quenched ensembles
and with meson masses above about
600\,MeV~\cite{Blum:2001xb,Noaki:2001un}. While these calculations
demonstrated the feasibility of the procedure, it is clear that,
in addition to using dynamical quarks it is necessary to perform
simulations at lighter masses and to go beyond the leading order
in the chiral expansion in order to understand the $\Delta I=1/2$
rule and the value of $\varepsilon^\prime/\varepsilon$. The
increase in computer power since 2001 and Next-to-Leading Order
(NLO) calculations in
$\chi$PT~\cite{Kambor:1989tz,Laiho:2002jq,Laiho:2003uy,Bijnens:1998mb,Lin:2002nq},
make it possible to contemplate such an endeavour. At NLO in
$\chi$PT, the number of LECs to be determined grows and it is not
possible to determine them all from $K\to\pi$ and $K\to$\,vacuum
matrix elements alone.

One possible approach to the determination of the NLO LECs is to
compute $K\to\pi\pi$ matrix elements. However, as explained in the
following section, there are difficulties in studying two-pion
states with isospin I=0. In this talk, we present an alternative
procedure for determining all the necessary LECs at NLO, based on
the evaluation of $K^0\pi^-\to\pi^-$ transitions (see
sec.\ref{sec:PiKPi}). A detailed analysis of this proposal will be
presented in a paper which is currently in preparation.

%%%%%%%%%%%%%%%%%%%%%%%%%%%%%%%%%%%%%%%%%%%%%%%%%%%%%%%%%%%%%%%%%%%%%%%%%%
%
% Section
%
%%%%%%%%%%%%%%%%%%%%%%%%%%%%%%%%%%%%%%%%%%%%%%%%%%%%%%%%%%%%%%%%%%%%%%%%%%
\section{Difficulties in the Evaluation of \boldmath{$K \to (\pi\pi)_{I=0}$}
Matrix Elements}
In this section we discuss the evaluation of $K\to\pi\pi$ matrix
elements and the difficulty encountered by the presence of
$s$-channel disconnected contractions. For illustration, consider
the process $K^0 \to \pi^+ \pi^-$, studied in $\chi$PT in
refs.\,\cite{Laiho:2002jq,Laiho:2003uy}. Using the operator
product expansion, the amplitude can be written in terms of matrix
elements of the $\Delta S=1$ weak effective Hamiltonian which
contains 10 four-quark operators $Q_i$ ($i=1$ -- 10):
\begin{equation}
\vev{ \pi\pi | \cH_{\Delta S=1} | K } = \frac{G_F}{\sqrt{2}}
\sum_{i=1}^{10}\, V^i_{CKM}\, c_i(\mu)\, \vev{ \pi\pi | Q_i | K
}_\mu\,,
\end{equation}
where $V_{CKM}^i$ are appropriate combinations of CKM matrix
elements, $c_i(\mu)$ are Wilson coefficients and $\mu$ is a
renormalization scale. To demonstrate the ideas let us consider
one of the operators:
\begin{equation}
Q_1 = \bar{s}_a \gamma_\mu ( 1 - \gamma_5) d_a ~ \bar{u}_b
\gamma^\mu ( 1 - \gamma_5) u_b .
\end{equation}
The correlation function from which the $K\to\pi\pi$ transition is
determined is
\begin{eqnarray}
C_{K\to\pi\pi}(t_K,t_O,t_\pi) &=& \vev{
\,0\,|\,\cO_{2\pi}(t_\pi)\,Q_1(t_O)\, \cO_K (t_K)\,|\,0\, }
\label{eq:corr_Kpipi} \\
& \approx & \vev{ \,0\, |\,  \cO_{2\pi}\, | \pi\pi }
            \vev{ \pi\pi| Q_1 | K }
            \vev{ K | \cO_K |\, 0\, } e^{ -E_{\pi\pi} ( t_\pi - t_O )}
                                  e^{ -E_K ( t_O - t_K )}, \label{eq:corr_Kpipi_asymp}
\end{eqnarray}
where in the last line we assume that the time intervals
$t_\pi-t_O$ and $t_O-t_K$ are sufficiently large and positive so
that contributions from heavier states can be neglected. $\cO_K$
and $\cO_{2\pi}$ are interpolating operators for the kaon and
two-pion states respectively. By computing $C_{K\to\pi\pi}$ one
can check the validity of expected asymptotic dependence on the
time intervals in eq.\,(\ref{eq:corr_Kpipi_asymp}) and extract the
matrix element $\vev{ \pi\pi| Q_1 | K }$.

%%%%%%%%%%%%%%%%%%%%%%%%%%%%%%%%%%%%%%%%%%%%%%%%%%%%%%%%%%%%%%%
\begin{figure}
\begin{center}
\begin{picture}(265,170)(-5,-105)
\Line(5,42)(40,42)\Line(50,40)(75,50.71)\Line(50,40)(75,29.29)
\Line(40,41)(75,56) \GOval(44,52)(12,3)(-20){1}
\Curve{(5,36)(75,24)}
\GCirc(40,40){5}{0.8}\Text(40,10)[c]{\small{\textbf{(A)}}}
\Text(79,57)[l]{{\small$\pi^+$}}\Text(1,40)[r]{{\small$K^0$}}
\Text(79,28)[l]{{\small$\pi^-$}}
\Line(175,42)(210,42)\Line(175,38)(210,38)\Line(210,41)(245,56)
\Line(210,39)(245,24)\Line(220,40)(245,50.71)
\Line(220,40)(245,29.29)
\GCirc(210,40){5}{0.8}\Text(210,10)[c]{\small{\textbf{(B)}}}
\Text(249,27)[l]{{\small$\pi^-$}} \Text(171,40)[r]{{\small$K^0$}}
\Text(249,57)[l]{{\small$\pi^+$}}
\Line(5,-58)(40,-58)\Line(5,-62)(40,-62) \Line(48,-60)(75,-44)
\Line(48,-60)(75,-76)
\Line(55,-60)(75,-48.15)\Line(55,-60)(75,-71.85)\GOval(44,-48)(12,3)(-20){1}
\GCirc(40,-60){5}{0.8}\Text(40,-100)[c]{\small{\textbf{(C)}}}
\Text(1,-60)[r]{{\small$K^0$}}\Text(79,-44)[l]{{\small$\pi^+$}}
\Text(79,-72)[l]{{\small$\pi^-$}}
\Line(175,-58)(210,-58)\Line(210,-57)(245,-42)\Line(210,-62)(245,-47)
\Line(210,-62)(245,-71)
\GCirc(210,-60){5}{0.8}\Curve{(175,-64)(245,-76)}
\Text(210,-100)[c]{\small{\textbf{(D)}}}
\Text(171,-60)[r]{{\small$K^0$}}\Text(249,-42)[l]{{\small$\pi^+$}}
\Text(249,-72)[l]{{\small$\pi^-$}}
\end{picture}\end{center}
\caption{Quark contraction diagrams for $K\to\pi\pi$ matrix
elements. The grey circle represents the insertion of a four-quark
operator. \label{fig:Diag1234}}
\end{figure}
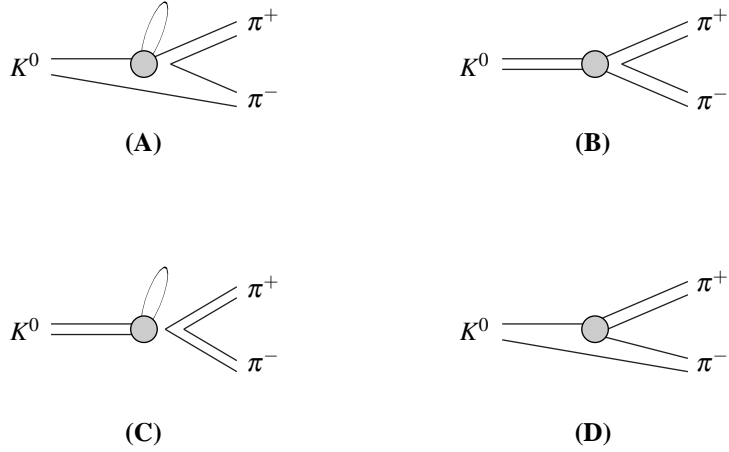
%%%%%%%%%%%%%%%%%%%%%%%%%%%%%%%%%%%%%%%%%%%%%%%%%%%%%%%%%%%%%%%

The quark flow diagrams which contribute to the correlation
function are sketched in Fig.\ref{fig:Diag1234}. The number of
quark propagators, and hence inversions of the Dirac operator,
required to evaluate these diagrams depends on the strategy as to
which of the times $t_K,\ t_{O}$ and $t_\pi$ are chosen for the
sources. If we choose to allow $t_\pi$ to vary, then we have to
solve the problem of inserting the quark propagator connecting
$\pi^+$ and $\pi^-$ in diagram (A,B,C) in Fig.\ref{fig:Diag1234}.
This requires as many quark propagators as the number of time
slices $t_{\pi}$. If instead we allow $t_O$ to vary, the insertion
of the loops in diagrams (A) and (C) require a large number of
inversions. If we fix both $t_O$ and $t_\pi$ then we are not able
to check the time behaviour of the two-pion state. Even if it were
possible to perform the large number of inversions which are
required to evaluate the diagrams, from experience we expect that
a very large number of configurations would be required to
evaluate the \textit{disconnected} diagram (C). Indeed one might
try evaluating the diagrams using stochastic all-to-all
propagators, but this may be noisier.

We stress that the difficulties described above are technical
rather than fundamental. They are nevertheless delaying the
evaluation of matrix elements with two-pion states in the I=0
channel (in the I=2 channel only diagram (D) needs to be
evaluated, which is relatively easy).

\begin{figure}
\begin{center}
\includegraphics[scale= .5]{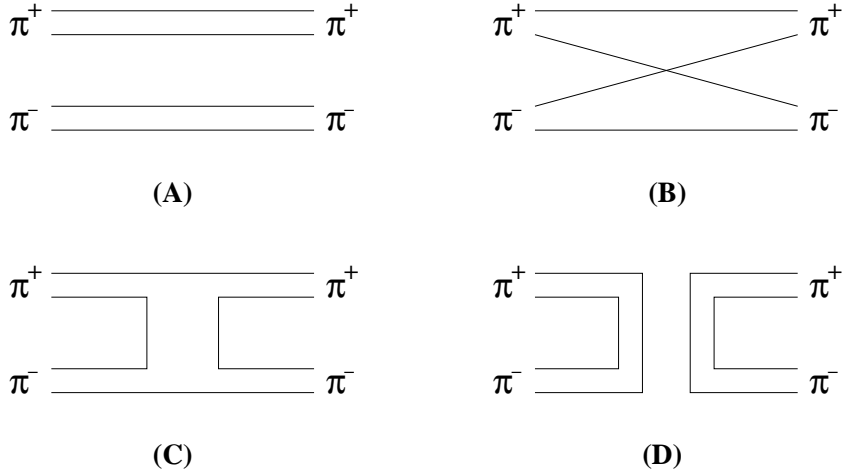}
\end{center}
\caption{Quark contraction diagrams for the propagator of the
$\pi^+\pi^-$ state. } \label{fig:DiagPiPi}
\end{figure}

In order to extract the matrix element $\vev{ \pi\pi| Q_1 | K }$
from $C_{K\to\pi\pi}$ we need to divide by the overlap factors,
including $\vev{0|\cO_{\pi\pi}|\pi\pi}$ (see
eq.\,(\ref{eq:corr_Kpipi_asymp})). The evaluation of
$\vev{0|\cO_{\pi\pi}|\pi\pi}$ requires the computation of the
propagator of the I=0 two-pion state, i.e. the evaluation of the
diagrams in Fig.\,\ref{fig:DiagPiPi}. This is at least as
challenging a task as that of the $C_{K\to\pi\pi}$, with similar
issues to the ones described above.

%%%%%%%%%%%%%%%%%%%%%%%%%%%%%%%%%%%%%%%%%%%%%%%%%%%%%%%%%%%%%%%
%%%%%%%%%%%%%%%%%%%%%%%%%%%%%%%%%%%%%%%%%%%%%%%%%%%%%%%%%%%%%%%

%%%%%%%%%%%%%%%%%%%%%%%%%%%%%%%%%%%%%%%%%%%%%%%%%%%%%%%%%%%%%%%
%
% section
%
%
%%%%%%%%%%%%%%%%%%%%%%%%%%%%%%%%%%%%%%%%%%%%%%%%%%%%%%%%%%%%%%%
\section{$K^0\pi^- \to \pi^-$ Matrix Elements}
\label{sec:PiKPi}

We start by stressing that our aim in this paper is limited to the
determination of the LECs at NLO in the chiral expansion, from
which we evaluate the $K\to\pi\pi$ decay amplitudes at the same
level of precision. The key point is that it is possible to
compute the LECs using any appropriate external states; in
particular it is not necessary to use two-pion states with $I=0$.
Our proposal is to determine the LECs for $\Delta S=1$ $\Delta
I=1/2$ operators by computing matrix elements for the (unphysical)
transition $K^0\pi^-\to\pi^-$ as well as $K\to\pi$ and
$K\to$\,vacuum matrix elements. In this section we explain the
advantages of this proposal; a detailed demonstration that all the
necessary LECs can be determined from these transitions for a
reasonable set of kinematic parameters will be presented in a
forthcoming publication. $K^0\pi^-$ is an $I=3/2$ highest weight
state, and, as we shall see below, this leads to a number of
important simplifications (analogous to those present when
studying the $I=2$ $\pi\pi$ state).

%%%%%%%%%%%%%%%%%%%%%%%%%%%%%%%%%%%%%%%%%%%%%%%%%%%%%%%%%%%%%%%
\begin{figure}
\begin{center}
\begin{picture}(265,170)(-5,-105)
\Curve{(5,12.18)(10,16.94)
(20,23.15)(30,27.32)(40,30.31)(60,33.89)(80,35)}
\Line(5,20)(40,40) \Line(5,60)(40,40)\GOval(50,40)(3,10)(0){1}
\Curve{(5,67.82)(10,63.06)(20,56.85)(40,49.69)(60,46.11)(80,45)}\GCirc(40,40){5}{0.8}
\Text(0,16)[r]{{\small$\pi^-$}} \Text(0,64)[r]{{\small$K^0$}}
\Text(85,42)[l]{{\small$\pi^-$}}
\Text(40,10)[c]{\small{\textbf{(A)}}}
\Curve{(175,12.18)(180,16.94)
(190,23.15)(200,27.32)(210,30.31)(230,33.89)(240,35)}
\Line(175,20)(210,40)\Line(175,58)(210,38) \Line(175,64)(210,44)
\Line(210,42)(240,42) \GCirc(210,40){5}{0.8}
\Text(170,16)[r]{{\small$\pi^-$}} \Text(170,64)[r]{{\small$K^0$}}
\Text(255,42)[l]{{\small$\pi^-$}}
\Text(210,10)[c]{\small{\textbf{(B)}}}
\Line(5,-42)(40,-62) \Line(5,-36)(40,-56)
\Curve{(5,-87.82)(10,-83.06)
(20,-76.85)(30,-72.68)(40,-69.69)(60,-66.11)(80,-65)}
\Curve{(5,-92.82)(10,-88.06)
(20,-81.85)(30,-77.68)(40,-74.69)(60,-71.11)(80,-70)}
\GOval(50,-60)(3,10)(0){1}
 \GCirc(40,-60){5}{0.8}
\Text(0,-92)[r]{{\small$\pi^-$}} \Text(0,-36)[r]{{\small$K^0$}}
\Text(85,-66)[l]{{\small$\pi^-$}}
\Text(40,-100)[c]{\small{\textbf{(C)}}}
\Curve{(175,-32.18)(180,-36.94)(190,-43.15)(210,-50.31)(230,-53.89)(240,-55)}
\Line(175,-42)(210,-60)\Line(210,-61)(240,-61)\Line(175,-78)(210,-58)
\Line(175,-84)(210,-64) \GCirc(210,-60){5}{0.8}
\Text(170,-81)[r]{{\small$\pi^-$}}
\Text(170,-36)[r]{{\small$K^0$}}
\Text(255,-56)[l]{{\small$\pi^-$}}
\Text(210,-100)[c]{\small{\textbf{(D)}}}
\end{picture}\end{center}
%\begin{center}
%\includegraphics[scale= .5]{newDiag1234.eps}
%\end{center}
\caption{Quark contraction diagrams for $K\pi\to\pi$ transitions.
The grey circle represents the insertion of a four-quark
operator.} \label{fig:newDiag1234}
\end{figure}
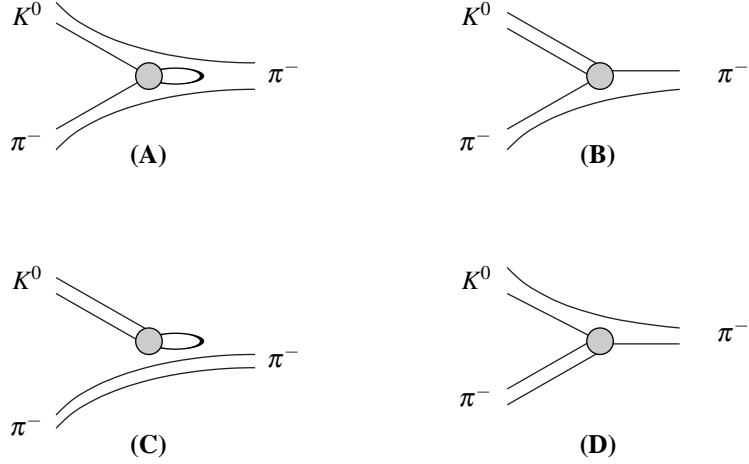
%%%%%%%%%%%%%%%%%%%%%%%%%%%%%%%%%%%%%%%%%%%%%%%%%%%%%%%%%%%%%%%

The correlation function corresponding to $K\pi\to\pi$ transitions
is
\begin{eqnarray}
C_{K^0\pi^-\to\pi^-}(t_K,t_O,t_\pi) &=& \vev{\,0\,|
\cO_{\pi^+}(t_\pi)Q_1(t_O) \cO_{\pi^-}(t_K) \cO_K (t_K)\,|\,0\, } \\
&&\hspace{-0.8in} \approx\vev{ 0 |  \cO_{\pi^+} | \pi^- }
            \vev{ \pi^-| Q_1 | K^0 \pi^- }
            \vev{ K^0 \pi^- |\, \cO_K \cO_{\pi^-}\, |\, 0\, }
                                  e^{ -E_{\pi} ( t_\pi - t_O )}
                                  e^{ -E_{K\pi} ( t_O - t_K )},
\end{eqnarray}
where $\cO_{\pi^-}$ and $\cO_{\pi^+}$ are interpolating operators
which can create or annihilate a $\pi^-$ respectively. The quark
contraction diagrams corresponding to this process are shown in
Fig.\ref{fig:newDiag1234}; they have common topologies with those
in Fig.\ref{fig:Diag1234}. The difference is that the final-state
$\pi^+$ in Fig.\ref{fig:Diag1234} is now \textit{crossed} into the
initial state where it is a $\pi^-$. Thus, the quark line
connecting $\pi^+$ to $\pi^-$ in diagrams (A,B,C) of
Fig.\ref{fig:Diag1234} now connects $\pi^-$ to $\pi^-$ propagating
from $t_K$ to $t_\pi$ in Fig.\ref{fig:newDiag1234}. We can
therefore vary $t_K$ using a small number of quark propagators,
and hence check the asymptotic exponential behavior of the $K\pi$
state.

What is more important is the absence of diagram such as (C) in
Fig.\ref{fig:Diag1234}. The $s$-channel disconnected diagram in
Fig.\ref{fig:Diag1234}(C) now becomes the $t$-channel disconnected
diagram in Fig.\ref{fig:newDiag1234}(C). Such $t$-channel
disconnected contributions appear, for example, when studying
$K\to\pi\pi$ matrix elements with an $I=2$ $\pi\pi$ state. There
have been several calculations of $I=2$ $\pi\pi$ states
\cite{Kim:2004sk,Yamazaki:2004qb} and this did not pose any
problem.

A related advantage is that one can avoid the vacuum subtraction.
For an $I=0$ final state at rest, there is mixing with the vacuum
state which must be subtracted~\footnote{One can avoid the vacuum
subtraction by considering the $\pi\pi$ state at non-zero
momentum}. In this case, the correlation function in
eq.\,(\ref{eq:corr_Kpipi}) contains the vacuum contribution,
\begin{equation}
 \vev{ 0 |  \cO_{2\pi} | 0 }
            \vev{ 0 | Q_1 | K^0 }
            \vev{ K^0 | \cO_K | 0 } e^{ -E_K ( t_O - t_K )}
\end{equation}
which must be subtracted in order to get
Eq.(\ref{eq:corr_Kpipi_asymp}). There is no such subtraction
necessary in the $K\pi\to\pi$ correlation function.

\begin{figure}
\begin{center}
\includegraphics[scale= .5]{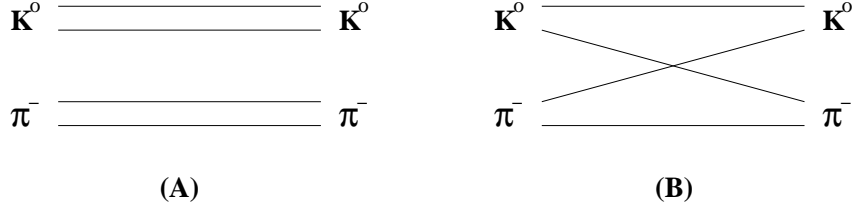}
\end{center}
\caption{Quark contraction diagrams for the propagator of the
$K^0\pi^-$ state.} \label{fig:DiagKPi}
\end{figure}

Note also that the overlap factor $\vev{K^0\pi^- | \cO_K
\cO_{\pi^-} | 0 }$ can be calculated much more easily.
Fig.\ref{fig:DiagKPi} shows the quark contraction diagrams. These
are the same diagrams which appear in the $I=2$ $\pi\pi$
propagator which is calculated without any difficulty.

When evaluating $K^0\pi^-\to\pi^-$ matrix elements, we need to
decompose the operators $Q_i$ into their $\Delta I=1/2$ and
$\Delta I=3/2$ components. In $K\to\pi\pi$ transitions, the
isospin transfer, $\Delta I$ is determined by isospin of the final
state, $\Delta I=1/2$ if $I_{\pi\pi}=0$ and $\Delta I=3/2$ if
$I_{\pi\pi}=2$. For the $K^0\pi^-\to\pi^-$ transition, the isospin
of the external states is already fixed, $I_{K^0\pi^-}=3/2$,
$I_{\pi^-} = 1$, both $\Delta I=1/2$ and $\Delta I=3/2$
transitions are allowed and therefore a basis of operators with
fixed isospin should be used (see for example Appendix A of
ref.\,\cite{Noaki:2001un}).

Since, for both $K\to\pi\pi$ and $K\pi\to\pi$ transitions we have
a two particle state, the finite-volume effects are not
exponentially small and should be taken into
account~\cite{Lellouch:2000pv}. Just as for the $(\pi\pi)_{I=2}$
state, the lowest energy $K^0\pi^-$ state(s) can be isolated and
the standard
techniques~\cite{Lellouch:2000pv,Kim:2005gf,Christ:2005gi} can be
applied. Energy, of course, must be injected at the operator
(which is also the case in practice in $K\to\pi\pi$ transitions).

The observation that the initial $K^0\pi^-$ state is one of
highest weight also means that one can also envisage performing
the $K\pi\to\pi$ calculations in the (non-unitary) partially
quenched QCD, which is not the case for the standard extraction of
$\Delta I=1/2$ $K\to\pi\pi$ matrix elements~\cite{Lin:2002aj}. One
might also try to improve the precision of the determination of
the LECs by extending the kinematical reach of the calculations
using partial twisting, which again is not possible for
$K\to(\pi\pi)_{I=0}$
transitions~\cite{Sachrajda:2004mi,Bedaque:2004ax}.

%%%%%%%%%%%%%%%%%%%%%%%%%%%%%%%%%%%%%%%%%%%%%%%%%%%%%%%%%%%%%%%
%%%%%%%%%%%%%%%%%%%%%%%%%%%%%%%%%%%%%%%%%%%%%%%%%%%%%%%%%%%%%%%

%%%%%%%%%%%%%%%%%%%%%%%%%%%%%%%%%%%%%%%%%%%%%%%%%%%%%%%%%%%%%%
%
% section
%
%%%%%%%%%%%%%%%%%%%%%%%%%%%%%%%%%%%%%%%%%%%%%%%%%%%%%%%%%%%%%%
\section{Concluding Remarks}\label{sec:discussion}

In this paper we have presented a suggestion for the determination
of the LECs necessary for the evaluation of the
$K\to(\pi\pi)_{I=0}$ decay amplitudes at NLO in the chiral
expansion. Our approach is based on the computation of
$K^0\pi^-\to\pi^-$ matrix elements (combined with $K\to\pi$ and
$K\to$\,vacuum matrix elements). We will demonstrate explicitly
that all the LECs can be determined in this way in a forthcoming
paper. We then have to develop effective strategies for
implementing these ideas in numerical simulations and to
investigate how precisely the LECs can be determined.

In spite of the relative simplicity of our approach, there will be
technical issues in its implementation; including the subtraction
of power divergences. As is well known, some of the $\Delta S=1$
four-quark operators mix with lower dimensional ones, leading to
ultra-violet divergences proportional to inverse powers of the
lattice spacing. The numerical subtraction of the divergences at
leading order in the chiral expansion was performed in the
quenched studies of refs.\,\cite{Blum:2001xb,Noaki:2001un} and it
remains to be seen how accurately this can be done in dynamical
simulations at NLO in the chiral expansion.

\end{document}